\input{aipcheck}

\documentclass[
    ,final            
  ]
  {aipproc}

\layoutstyle{6x9}

\begin{document}

\title{Mechanisms of exclusive meson production \\ at high energies}


\author{Antoni Szczurek}{
  address={Institute of Nuclear Physics PAN, PL-31-342 Cracow, Poland\\
       University of Rzesz\'ow, PL-35-959 Rzesz\'ow, Poland}
}



\begin{abstract}
I discuss mechanisms of exclusive production
of mesons at high energies. Some illustrative examples
for the FNAL Tevatron and CERN LHC as well as for lower
energies are shown.
\end{abstract}

\maketitle


\section{Introduction}

\begin{figure}
  \includegraphics[height=.25\textheight]{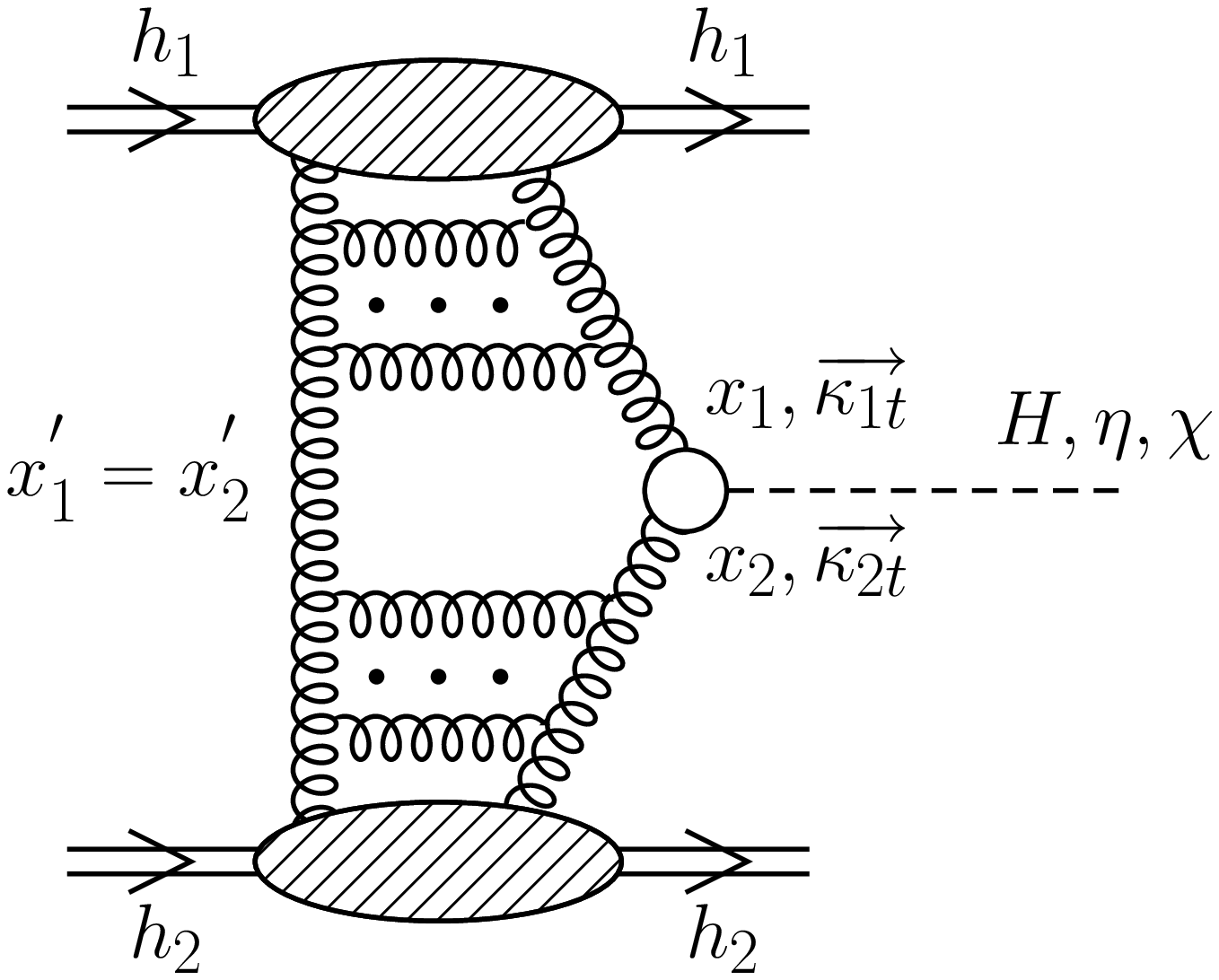}
  \caption{A sketch of the bare QCD mechanism of
exclusive heavy scalar meson production.
\label{fig:scalar_diagram}
}
\end{figure}

\begin{figure}
  \includegraphics[height=.25\textheight]{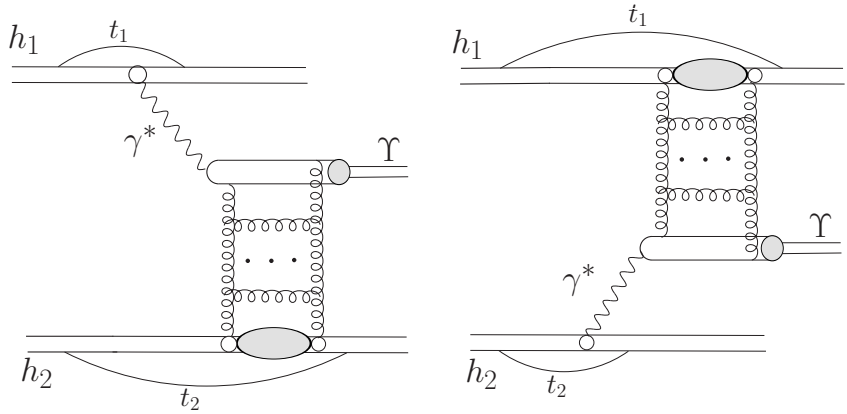}
  \caption{Two basic QED $\otimes$ QCD mechanisms of
exclusive heavy vector meson production.
\label{fig:vector_diagram}
}
\end{figure}

\begin{figure}
  \includegraphics[height=.25\textheight]{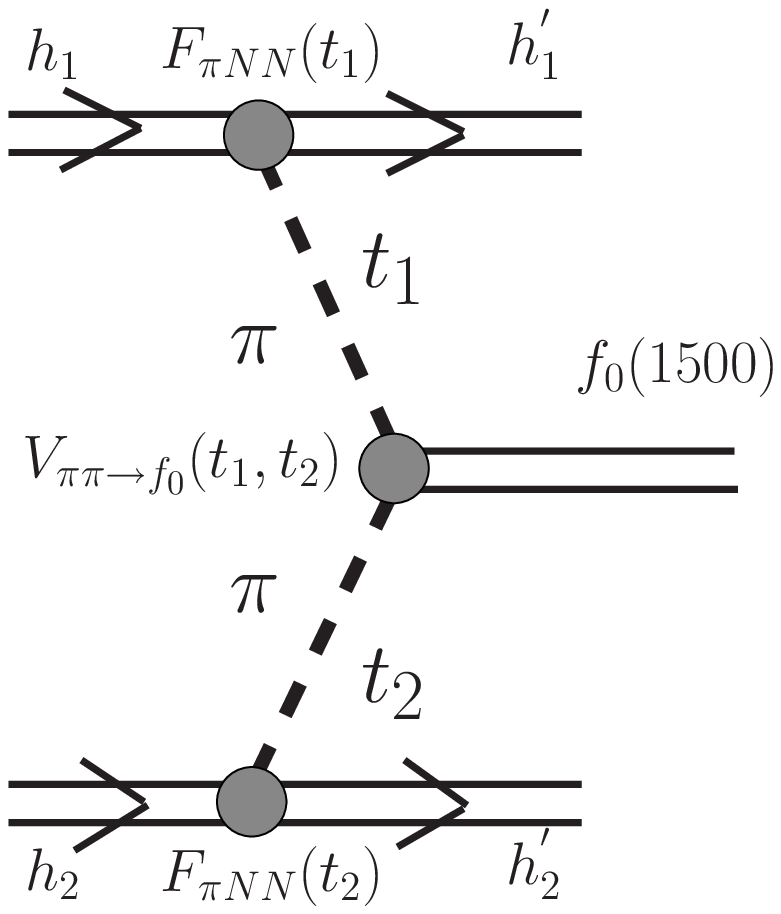}
  \caption{A sketch of the bare QCD mechanism of
exclusive heavy scalar meson production.
\label{fig:pion_pion_diagram}
}
\end{figure}

The exclusive production of mesons was studied in the past
mostly close to the kinematical threshold.
The Tevatron is a first accelerator which opens
a possibility to study the (semi)exclusive production
of mesons at high energies. A similar program will be
carried out in the future at just being put into operation
LHC.
Here I briefly review several mechanisms of exclusive
meson production studied recently by the Cracow group
(the details can be found in \cite{SPT07,SS07,PST08,RSS08,SL08}).
In general, the mechanism of the reaction depends
on the quantum numbers of the meson and/or its internal
structure.
For heavy scalar mesons 
(scalar quarkonia, scalar glueballs) the mechanism of 
the production, shown in Fig.\ref{fig:scalar_diagram}, 
is exactly the same as for the diffractive Higgs boson 
production extensively discussed in recent years
by the Durham group \cite{Durham}.
The dominant mechanism for the exclusive heavy 
vector meson production is quite different.
Here there are two dominant processes shown in 
Fig.\ref{fig:vector_diagram}. When going to lower
energies the mechanism of the meson production becoming
more complicated and usually there exist more mechanisms. 
For illustration in Fig.\ref{fig:pion_pion_diagram} 
I show a new mechanism of the glueball production 
proposed recently in Ref.\cite{SL08}.

\section{Some examples}

Recently we have calculated differential cross sections 
in several reactions (including different mechanisms):
\begin{itemize}
\item $ p p \to p p \eta' $, $ p p \to p p \eta_c$
 ( IP IP + $\gamma \gamma$ ) 
\item $ p p \to p p \chi_c(0)$ (IP IP + $\gamma \gamma$ ) 
\item $ p p \to p p f_0(1500)$ (IP IP + $\pi^+ \pi^-$)
\item $ p p \to p p J/\psi$ (IP $\gamma$ + $\gamma$ IP)
\item $ p p \to p p \Upsilon$ ( IP $\gamma$ + $\gamma$ IP)
\item $ p p \to p p \pi^+ \pi^-$ ( (IP + IR) $\otimes$ (IP + IR))
\end{itemize}
The details of the formalism as well as a detailed analysis
of differential distribution in longitudinal and transverse
momenta can be found in the original papers
\cite{SPT07,SS07,PST08,RSS08,SL08}).
Here I wish to flash only some illustrative examples.

\subsection{Scalar meson production}

\begin{figure}
  \includegraphics[height=.3\textheight]{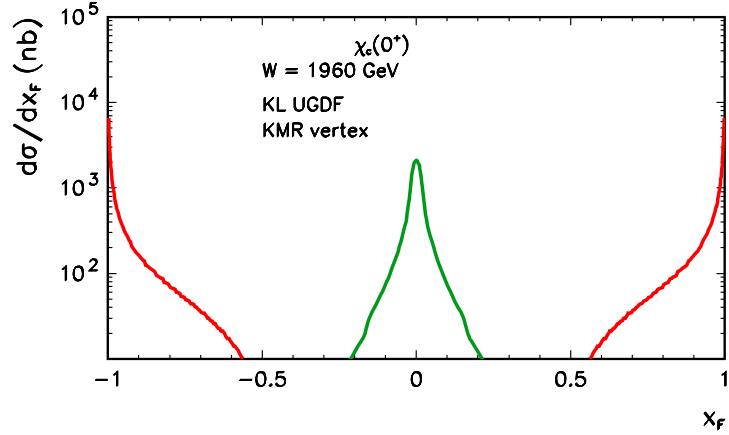}
  \caption{Distribution in Feynman $x_F$ of $\chi_c(0)$ 
meson and associated proton (left bump) and antiproton (right bump) for the Tevatron energy.
\label{fig:dsig_dxi_kl}}
\end{figure}

Let us start with the production of scalar particles.
In Ref.\cite{PST08} we discussed in detail the production
of scalar $\chi_c(0)$ meson. Here the dominant mechanism
is exactly the same as for celebrated recently
diffractive production of Higgs boson.
A study of this reaction can be an alternative
for inclusive searches for the Higgs boson at LHC. 
In order to show the general features of the exclusive 
diffractive production in Fig.\ref{fig:dsig_dxi_kl} I show
distribution in Feynman $x_F$ of $\chi_c(0^+)$ mesons
(the middle bump) as well as distributions of the 
associated proton (left bump) and antiproton (right bump)
for W = 1960 GeV.
In this calculation the unintegrated gluon distribution
(UGDF) ala Kharzeev-Levin were used \cite{PST08}.
As discussed in Ref.\cite{PST08} the cross section strongly
depends on UGDF used. 
A clear gaps in $x_F$ between the meson
and associated nucleons can be seen. The gaps in $x_F$
translate into gaps in rapidity. Generally the diffractive
production of single mesons can be characterized by
rapidity gaps. Can this be used as a criterion for 
selecting appropriate events?
It would be useful for potential experiments to calculate 
cross section for inclusive double diffraction of 
$\chi_c(0)$ mesons to verify if the large rapidity gap
criterion can be used to pin down the exclusive channel.

It seems that the dominant mechanism of the glueball
production at high energies should be the same as for
the $\chi_c(0)$ meson. If there is appreciable gluonic
component in a meson the ladder gluons
could (should) strongly couple to the meson.
In Ref.\cite{SL08} we concentrated rather on the 
intermediate and low energy regime. In addition to the QCD
diffractive mechanism we considered the two-pion fusion.
It seems that this new mechanism may dominate close to 
threshold. This can be checked in the future with
the PANDA detector at the complex FAIR planned at 
GSI Darmstadt. In the $p \bar p \to p \bar p f_0(1500)$ 
channel the pionic mechanism dominates close to threshold,
while the QCD diffraction takes over at larger energies.
In the $p \bar p \to n \bar n f_0(1500)$
channel the pion-exchange mechanism may dominate
in a broader range of energies.

\begin{figure}
  \includegraphics[height=.3\textheight]{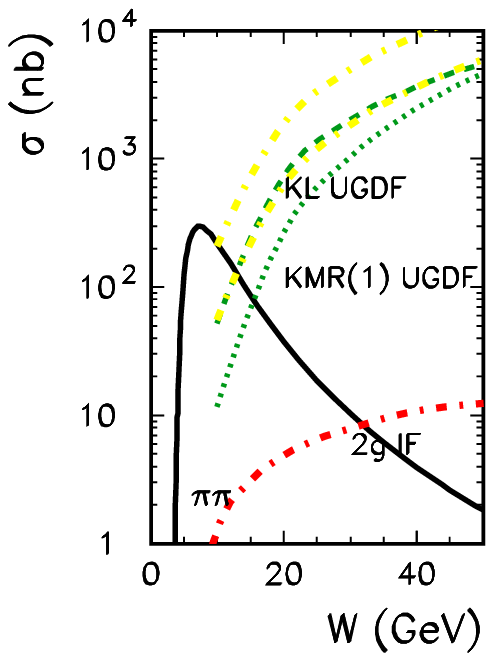}
  \includegraphics[height=.3\textheight]{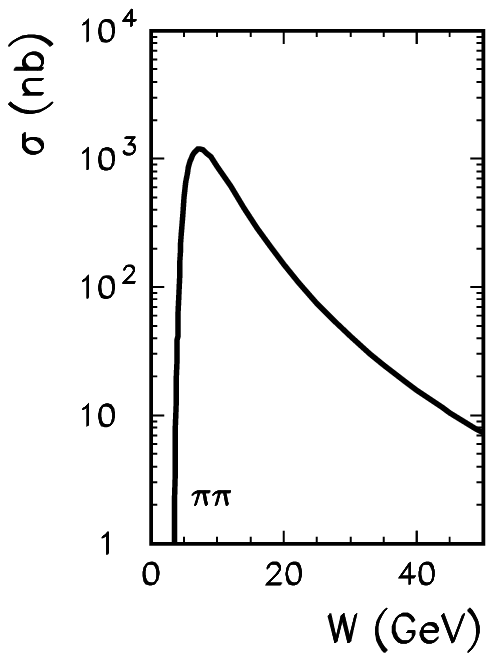}
  \caption{Integrated cross section as a function of initial energy. The solid line corresponds to the pion-pion fusion whereas the dashed and dotted line to the diffractive QCDmechanism with different UGDFs.}
\end{figure}

\subsection{Vector meson production}

\begin{figure}
  \includegraphics[height=.3\textheight]{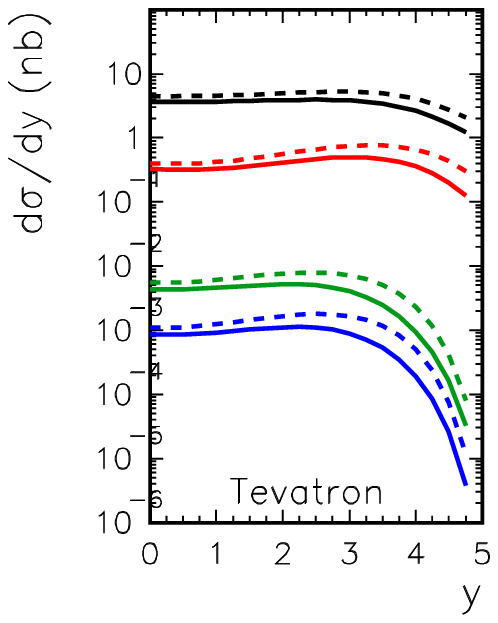}
  \includegraphics[height=.3\textheight]{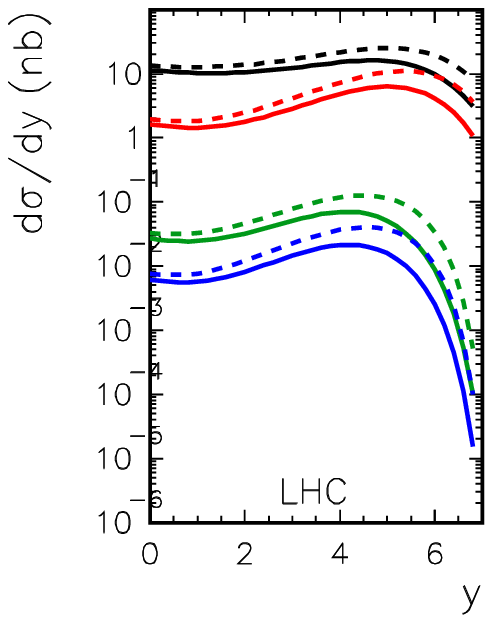}
  \caption{Distribution in rapidity of
$J/\Psi$, $\Psi'$, $\Upsilon$, $\Upsilon'$ (from top to bottom) for Tevatron (left panel) and LHC (right panel).
The dashed line corresponds to calculation in 
the Born approximation and the solid line includes 
absorption corrections.
\label{fig:dsig_dy_vector}}
\end{figure}

Because of their quantum numbers vector mesons cannot be
produced in a fusion of two pomerons (two gluonic ladders).
For heavy vector quarkonia the simplest possible mechanism
is a photon-pomeron or pomeron-photon fusion shown 
in Fig.\ref{fig:vector_diagram}. For light vector mesons 
the situation is slightly more complicated.
In Ref.\cite{SS07} we discussed several differential
distributions for exclusive $J/\Psi$ production
in a phenomenological model. In Ref.\cite{RSS08}
we performed a similar analysis for exclusive production
of $\Upsilon$ in the formalism of unintegrated
gluon distributions. In Fig. \ref{fig:dsig_dy_vector}
I show a compilation of the rapidity distributions
for $J/\Psi$, $\Psi'$, $\Upsilon$ and $\Upsilon'$
for both Tevatron (left panel) and LHC (right panel). 
Both distributions obtained with bare amplitudes 
(dashed lines) and 
including absorption corrections (solid lines) are
shown. 
More details about absorption corrections can be found 
in Refs.\cite{SS07,RSS08}.
The transverse-momentum integrated rapidity distribution
is only mildly modified by the absorption corrections.
Much bigger effects can be seen at large transverse
momenta \cite{SS07,RSS08} where the cross section is 
much smaller. We hope that in a near future the CDF
collaboration at the Tevatron will release the experimental
cross section for exclusive production of quarkonia
\cite{Albrow}.

\section{Conclusions}

I have briefly discussed some of our results
on exclusive production of mesons at high energies.
At present a direct comparison with experimental
data is not possible.
We expect some experimental data from the Tevatron soon.
In a more distant future one may hope for experimental 
data from LHC.

\begin{theacknowledgments}
The collaboration with Wolfgang Sch\"afer, Roman Pasechnik,
Oleg Teryaev, Anna Cisek and Piotr Lebiedowicz
on the topics presented here is acknowledged.
\end{theacknowledgments}

\end{document}